# Deterministic multidimensional growth model for small-world networks

Aoyuan Peng, Lianming Zhang[*]

*College of Physics and Information Science, Hunan Normal University, Changsha, 410081, China
Institute of Physics and Key Laboratory of Low Dimensional Quantum Structures and Quantum Control of Ministry of Education, Hunan Normal University, Changsha, 410081, China*

**Abstract**: We proposed a deterministic multidimensional growth model for small-world networks. The model can characterize the distinguishing properties of many real-life networks with geometric space structure. Our results show the model possesses small-world effect: larger clustering coefficient and smaller characteristic path length. We also obtain some accurate results for its properties including degree distribution, clustering coefficient and network diameter and discuss them. It is also worth noting that we get an accurate analytical expression for calculating the characteristic path length. We verify numerically and experimentally these main features.

## 1 Introduction

In previous decades, there has been great interest in small-world effect existing widely in real-world complex systems [1], such as World Wide Web, food webs, scientific-collaboration networks, airport networks and social networks which are shown to be small-world networks. Small-world networks have the characteristics from both regular and random networks. These small-world networks are characterized by the following two distinguishing properties: (i) local clustering coefficient $C$ is large, similar to a regular lattice, and (ii) characteristic path length $L$ is small, growing as $\ln(N)$ or shower, similar to a random graph. Coexistence of high local clustering and small characteristic path length of networks is called the small-world effect. The long-range connections between nodes of the complex networks with the small-world effect play an important role in enhancing the speed of epidemic spreading and signal propagation, improve the efficiency of network congestion control, resource search and information navigability.

Watts and Strogatz [2] published, randomly rewiring each edge of a ring lattice with some probability, seminal work on how small-world networks are modeled. It has induced an avalanche of works aiming at uncovering the small-world effect of complex networks and setting up related models to expound the generative mechanism of small-world effect. Since then many more variants and generalizations of small-world network models have been studied [3].

Of particular interest are two classes of studies. The first-class, in which the models for small-world networks are stochastic and modeled by adding randomness to regular graphs. For example, Newman and Watts [4] adapted Watts and Strogatz model by taking the place of the rewiring phase using just addition of long-range connections between randomly chosen nodes on the ring lattice with some probability. Jespersen and Blumen [5] proposed a general model for small-world networks by adding extra edges to a ring lattice with the probability as a function of their mutual chemical distance. The edge reconnection is governed by a function that depends on the distance between the nodes in small-world network model proposed by Kuperman and Abramson [6]. The long-range connections are added to a *d*-dimensional lattice controlled by a clustering exponent that controls the probability of an edge between nodes as a function of their lattice distance in small-world network model proposed by Kleinberg [7]. Guo et al. [8] proposed a simple model for small-world networks by two-stage adding process for each new node. Most research on these stochastic models for small-world networks has used probabilistic techniques, random replacements and additions of some edges of a regular graph to research the limiting properties of these networks. The randomness of stochastic small-world network models makes it more difficult to understand on how do different nodes of real complex networks interact with each other. It also makes is more difficult to characterize on how do the size (the number of nodes and the number of edges) of real complex networks increase over time [9].

---

[*] Corresponding author.
E-mail addresses: lianmingzhang@gmail.com (L. Zhang)

The second-class, in which the models for small-world networks can be modeled in a deterministic way. The deterministic models can conquer the disadvantage of stochastic small-world network models. Comellas et al. [10] [11] introduced a deterministic small-world model for complex networks as communication networks using graph-theoretic methods to get accurate results, and presented a recursive graph construction which produces small-world networks [12]. Ozik et al. [13] proposed a deterministic growth model for small-world networks by beginning with connected nodes on the ring lattice. They placed a new node in a randomly chosen inter-node interval along the ring lattice and connect the new node to all of its nearest neighbors. Zhang et al. [14] modified Ozik model, which can unify a regular ring lattice and Ozik model to the same framework by tuning a parameter. They also created a variant of Ozik model by beginning with a triangle using edge iterations [9]. Xiao and Parhami [15] proposed a simple model for deterministic small-world networks based on Cayley graphs. In addition, the deterministic models for complex networks with small-world effect and hierarchical structure [16] [17], mixing small-world effect and scale-free feature [18], attracts much attention.

The existing deterministic small-world network models are almost based on the simple planar lattice. They make it more easy to understand the interacting between nodes, but do it more difficult to apply to characterize the properties of the complex networks with geometric space structure.

In this paper, we present a deterministic multidimensional growth (DMG) model for small-world networks with geometric space structure. Compared with probabilistic methods, an advantage of the DMG model is that we can model some particular networks with small-world effect if one of the important properties of network topology is given. These properties include average degree, degree distribution, clustering coefficient, characteristic path length and network diameter. Most importantly, the DMG model can characterize the geometric space structure and the main characteristics of complex networks with small-world effect. In addition, we use a new simple method to obtain the accurate analytical expression for calculating the characteristic path length of the DMG model, and we verify numerically and experimentally these main features.

The rest of the paper is organized as follows. In Section 2 we propose a deterministic multidimensional growth model for small-world networks with geometric space structure. In Section 3 we analyze the main properties of the DMG model and obtain related accurate expressions for calculating them as a function of the iteration steps as well as the network size. Section 4 contains the summary of results and some concluding remarks.

## 2 The DMG model

In this section, we introduce the DMG model for small-world networks with geometric space structure in deterministic way, and we denote this network graph by DMG($t$) after $t$ iteration steps. The construction algorithm of the DMG model is the following: (i) for $t = 0$, there are four early nodes, and the four nodes connect with one another and they form a connected graph (triangular pyramid), denoted by DMG(0); (ii) for $t \geq 1$, the graph, denoted by DMG($t$), get from the graph DMG($t-1$), created at iteration step $t-1$, by adding a new node for each edge of the graph DMG($t-1$) and attaching it to two end nodes of the edge. The first four steps of the generation process of the DMG model for small-world networks is shown in Figure 1.

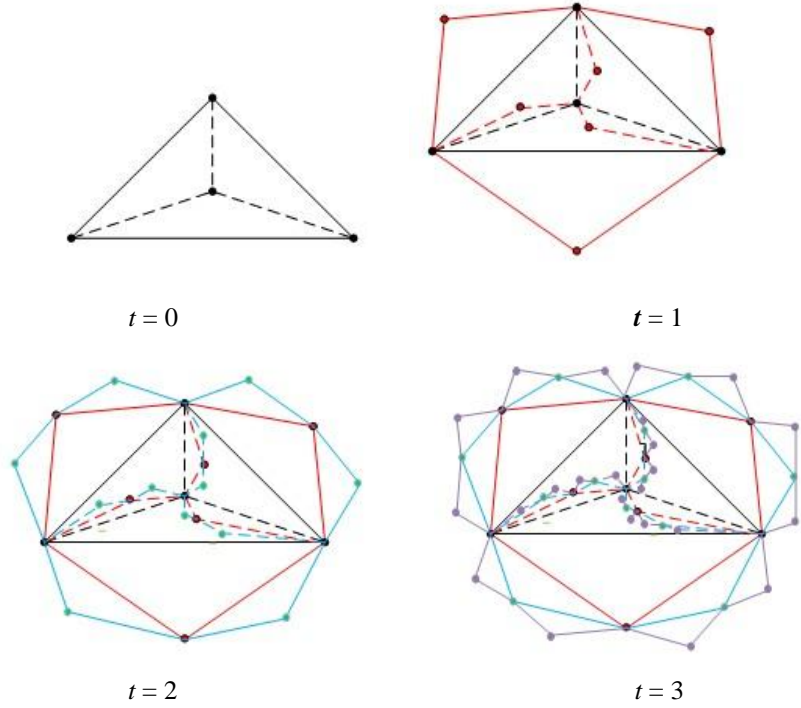

$t = 0$      $t = 1$

$t = 2$      $t = 3$

**Figure 1. The generation process of the DMG model for small-world networks**

For $t = 0$, we hold the number of edges $e_0 = 6$, and the number of nodes $n_0 = 4$ in the graph DMG(0). Without loss of generality, the total number of the nodes is denoted by $n_t$ and the total number of the edges in the graph DMG($t$) is denoted by $e_t$ for $t>0$.

Obviously, for $t > 0$, the number of the new nodes in the graph DMG($t + 1$) is twice that of the new nodes in the graph DMG($t$). The latter is also twice that of the new nodes in the graph DMG($t$). There are $\Delta n_{t+1} = 2 \times \Delta n_t$, $\Delta e_t = 2 \times \Delta n_t$ and $\Delta n_1 = 6$. Therefore, we hold

$$\Delta n_t = 3 \times 2^t, \; t \geq 1 \tag{1}$$

and

$$n_t = n_0 + \Delta n_1 + \Delta n_2 + \ldots + \Delta n_t = 3 \times 2^{t+1} - 2 \tag{2}$$

and then

$$e_t = 3 \times 2^{t+2} - 6 \tag{3}$$

From Equation (2) and Equation (3), the network size (the total number of nodes) and the total number of edges increase exponentially with iteration step *t*, and it is helpful to the fast generation of large-scale network topology.

## 3 Characteristics of the DMG model

In the following, we derive the analytical solutions and present the numerical and simulation results of some main topology properties, such as the degree distribution, the clustering coefficient, the network diameter and the characteristic path length, after then we analyse and compare the results.

*3.1. Degree distribution*

The degree of a node (node degree) is the number of edges connected to the node. It is also the number of its direct neighbors. A node is more important for larger value of the degree. The average degree of networks is mean value of the degrees of all nodes in the network, and the connectivity of networks is more good for larger value of the average degree. The degree distribution can characterize the topology structure of networks. For example, the degree distribution of regular lattices is Delta distribution, and that of complete random graphs is approximate Poisson distribution, and then that of scale-free networks is power-law distribution.

According to the average degree definition, Equation (2) and Equation (3), we can derive the average degree of the graph DMG(*t*), as follows

$$<k>(t) = \frac{2e_t}{n_t} = 4 - \frac{2}{3 \times 2^t - 1} \tag{4}$$

Form Equation (4), we know the average degree is infinite close to 4 for $t \to \infty$. This means that, in large-scale small-world networks, the direct neighbors of each node is 4 on average, and it's obviously a sparse graph.

The degree $k_0(0)$ of four nodes, denoted $n_0$, in the graph DMG(0) is 3. The degrees of four nodes $n_0$ are equal for the graph DMG(*t*) at the same iteration step *t*. The degree of the four nodes $n_0$ satisfies $k_0(t) = 3 + 3t$.

Besides the four nodes $n_0$, the degree of the node *i* is denoted $k_i(t)$ in the graph DMG(*t*), and the degree of the node *i* is denoted $k_i(t + 1)$ in the graph DMG(*t* + 1). Here, we have $k_i(t + 1) = k_i(t) + 2$. In the graph DMG(*t*), the degree of the node *i*, adding in the graph DMG(1), satisfies $k_i(t) = 2t$.

Now, let we assume the node *A* is created in the graph DMG($t_A$). We have $k_A(t_A) = 2$ and $k_A(t) = 2(t + 1 - t_A)$, where $t \geq t_A$. Obviously, the degree sequence of all nodes in the graph DMG(*t*) is $2 \times 1, 2 \times 2, 2 \times 3, 2 \times 4, \ldots, 2 \times t$ and $3 + 3t$. Let we assume the node with *k*-degree is created in the graph DMG($t_k$), and we have

$$t_k = t - k/2 + 1 \tag{5}$$

Cumulative degree distribution function *P*(*k*) indicates the proportion of the nodes whose degree is more than *k*, and that is $P(k) = P(k' > k)$[19]. The minimum degree of the DMG model is 2, so the cumulative degree distribution probability of the nodes with 2-degree is 1. The degree of the four early nodes is always the largest, and it means that they are important in the DMG model. From defining the accumulative degree distribution function and Equation (5), we can obtain

$$P(k) = P(k' > k) = P(t < t_k) = \frac{n_{t_{k-1}}}{n_t} = \frac{3 \times 2^{t+1-\frac{k}{2}} - 2}{3 \times 2^{t+1} - 2} = \frac{2^{-\frac{k}{2}}}{1 - 1/(3 \times 2^t)} - \frac{1}{3 \times 2^t - 1} \quad (6)$$

So, we can derive the degree distribution function $p(k)$ satisfies

$$p(k) = P(k' > k - 1) - P(k' > k) = \frac{(\sqrt{2} - 1) \times 2^{-\frac{k}{2}}}{1 - 1/(3 \times 2^t)} \quad (7)$$

For $t \to \infty$, we have $P(k) \approx 2^{-k/2}$ and $p(k) \approx (\sqrt{2} - 1) \times 2^{-k/2}$. For the large-scale networks, $P(k)$ and $p(k)$ are similar to the results in [7], and they possess the same index function distribution.

### *3.2. Clustering coefficient*

The real complex networks, such as World Wide Web, Internet, food webs and cell nets, have high local clustering, and WS model [2] first reunites the clustering feature of complex networks and the necessity of random graphs. The DMG model for small-world networks, proposed in this paper, may connects the clustering feature of complex networks with geometric space structure and the necessity of deterministic networks. Clustering coefficient describes the clustering feature. Clustering coefficient can classify local clustering coefficient, denoted $c_i$, and global clustering coefficient, denoted $C$. The local clustering coefficient $c_i$ of a node $i$ is the ratio of the total number $E_i$ of edges, which exist among it and its direct neighbors, and the number $k_i(k_i - 1)/2$ of all possible edges between them. The global clustering coefficient $C$ is the average value of the local clustering coefficient of all nodes.

Obviously, the degrees of the four early nodes $n_0$ are 3 in the graph DMG(0). Adding a node, the degree of $n_0$ and the edges between $n_0$ and its direct neighbors both increase 1, the early node degree increase 1, and the neighbor node connected edge also increase 1. The degree of the four early nodes is equal to the number of the edge between it and its neighbors, and the local clustering coefficient $c_0$ of the four early nodes in the DMG($t$) can be expressed as

$$c_0 = \frac{k_i}{C_{k_i}^2} = \frac{3 + 3t}{(3 + 3t)(3 + 3t - 1)/2} = \frac{2}{3t + 2} \quad (8)$$

The degree of a new node $i$ is 2 and the number of the edges, connected with its neighbor nodes, is 1. The degree of the node $i$ and the number $E_i$ of edges both increase 1 at each iteration step. We have $E_i = K_i - 1$ at all steps. The number of all possible edges between the node $i$ and its direct neighbors is $C_{k_i}^2$. The local clustering coefficient $C_i$ of the nodes $i$ as follows

$$C_i = \frac{E_i'}{C_{k_i}^2} = \frac{K_i - 1}{K_i(K_i - 1)/2} = \frac{2}{K_i} = 1/t \quad (9)$$

Clearly, the degree sequence of all the nodes in the network is $2 \times 1, 2 \times 2, \ldots, 2 \times t$ and $3 + 3t$, From Equation (8) and Equation (9), the local clustering coefficient $c_i$ is 1, 1/2, …, 1/t and 2/(3t + 2), respectively. The number of the nodes with the same degree is $\Delta n_t, \Delta n_{t-1}, \ldots, \Delta n_1$ and 4. So the global clustering coefficient $C$ of the network satisfies

$$
\begin{aligned}
C &= \frac{1}{n_t}[\sum_{i=1}^{t}\frac{1}{i}\times \Delta n(t-i+1)+\frac{2}{3t+2}\times 4] \\
&= \frac{1}{3\times 2^{t+1}-2}[1\times 3\times 2^{t}+\frac{1}{2}\times 3\times 2^{t-1}+\cdots+\frac{1}{t}\times 3\times 2^{1}+4\times \frac{2}{3t+2}] \\
&= \frac{1}{1-\frac{1}{3\times 2^{t}}}\times \ln 2+\frac{4}{3\times 2^{t+1}-2}\times \frac{2}{3t+2}
\end{aligned}
\qquad (10)
$$

For $t \to \infty$, the global clustering coefficient $C \approx \ln(2) \approx 0.693$, and it coincides with the result given in [7]. Figure 2 shows the impact of the iteration steps and the network size on the global clustering coefficient. Obviously, the global clustering coefficient increases with the increase of the number of iterative steps and converges to a stable value of 0.693. For example, the global clustering coefficient has reached 0.693 if iteration step is 8, and the number of nodes in the network is 1534 and the number of edges is 19355512. That means the large-scale networks, for example for $n > 1534$, created by the DMG model can ensure the global clustering coefficient remains almost the same.

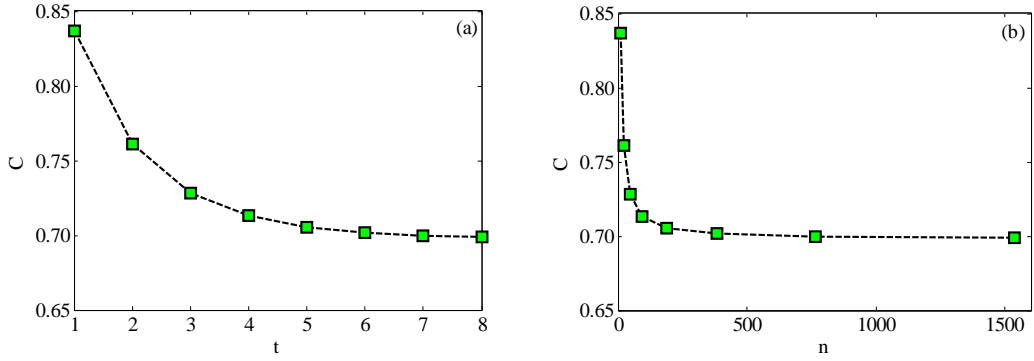

**Figure 2. The global clustering coefficient of the DMG model for small-world networks.** (**a**) The global clustering coefficient $C$ versus the iteration steps $t$; (**b**) The clustering global coefficient $C$ versus the network size $n$

### *3.3. Diameter and characteristic path length*

Using network diameter or characteristic path length, we can measure the transmission performance and the efficiency of networks. The former is the longest distance between two nodes in the network. For the DMG model, if $t = 0$, the network diameter $D = 1$, and if $t = 1$, $D = 3$. If $t \geq 2$, the network diameter always lies between a pair of nodes that have just been created at this step [9]. If $t$ is even, the network diameter $D = t + 1$, or else the network diameter $D = t + 2$.

Characteristic path length is the average value of the distances (the shortest path length) between every two nodes in the network, and it is much more difficult to calculate than connectivity. Many previous research on approximate calculation of the characteristic path length of small-world networks has based on the renormalization group method or the mean field method. Now, for random small-world network models, there are not accurate and analytical expressions of the characteristic path length. For some deterministic small-world network models, there are some accurate solutions using various methods. In this paper, we take over the method introduced in [20].

The characteristic path length of the graph DMG($t$) can be expressed as

$$L(t) = \frac{1}{n_t(n_t - 1)} g(t) \qquad (11)$$

where $n_t$ is given by Equation (2), and $g(t) = d(n_t, n_t)$, $d(X, Y) = \sum_{u \in X} \sum_{v \in Y} d(u, v)$.

A node $i \in \Delta n_t$ always connects two nodes, one of which is created at the iteration step $t - 1$ and the other one is created before the iteration step $t - 1$. If the node $i \in \Delta n_t$, the node $j \in \Delta n_{t-1}$ and then the node $l \in \Delta n_{t-2}$. If $j$ and $l$ are the two direct neighbors of $i$, we say $j$ the mother of $i$, and do $l$ the father of $i$. The two nodes, having the same mother, are brothers.

There are, by definition, $n_t = n_{t-1} \cup \Delta n_t$, so

$$d(n_t, n_t) = d(n_{t-1}, n_{t-1}) + d(\Delta n_t, n_{t-1}) + d(n_{t-1}, \Delta n_t) + d(\Delta n_t, \Delta n_t) \qquad (12)$$

For an undirected graph, we have $d(\Delta n_t, n_{t-1}) = d(n_{t-1}, \Delta n_t)$, and then

$$g(t) = g(t-1) + 2d(\Delta n_t, n_{t-1}) + d(\Delta n_t, \Delta n_t), \ t \geq 2 \qquad (13)$$

In fact, we can obtain $g(0) = 12$, $g(1) = g(0) + 2d(\Delta n_1, n_0) + d(\Delta n_1, \Delta n_1) = 12 + 2 \times 6 \times (1 + 1 + 2 + 2) + 6 \times (2 + 2 + 2 + 2 + 3) = 150$, $g(2) = g(1) + 2d(\Delta n_2, n_1) + d(\Delta n_2, \Delta n_2) = 150 + 2 \times 12 \times 21 + 12 \times 30 = 1014$.

First, we calculate $d(\Delta n_t, n_{t-1})$ for $t \geq 2$ in Equation (13). For $t \geq 2$, a new node, by construction algorithm, always connects with one father node and one mother node. The most short path between the new node and other nodes, not including its brother nodes, passes through its father node. The shortest distance between brother nodes is 2, but the distance through father node is 3. So, here we need a correcting factor $-|\Delta n_t|$. One of the four early nodes connects new three nodes, and each father node connects new two nodes. Without loss of generality, we handle one of new three nodes connected with the four early nodes alone, then

$$d(\Delta n_t, n_{t-1}) = 2d(n_{t-2}, n_{t-1}) + (|\Delta n_t| - 4) |n_{t-1}| - |\Delta n_t| + d(n_0, n_{t-1}) + 4|n_{t-1}|$$
$$= 2d(n_{t-2}, n_{t-1}) + |\Delta n_t| |n_{t-1}| - |\Delta n_t| + d(n_0, n_{t-1}), \ t \geq 2 \qquad (14)$$

where

$$d(n_{t-2}, n_{t-1}) = d(n_{t-2}, n_{t-2}) + d(n_{t-2}, \Delta n_t) = g(t-2) + d(n_{t-2}, \Delta n_t) \qquad (15)$$

From Equation (13), we have

$$2d(\Delta n_{t-1}, n_{t-2}) = g(t-1) - g(t-2) - d(\Delta n_{t-1}, n_{t-1}) \qquad (16)$$

Substituting Equations (15) and (16) into Equation (14), we have

$$d(\Delta n_t, n_{t-1}) = g(t-1) + g(t-2) - d(\Delta n_{t-1}, \Delta n_{t-1}) + |\Delta n_t| |n_{t-1}| - |\Delta n_t| + d(n_0, n_{t-1}) \qquad (17)$$

Next, we calculate $d(\Delta n_t, \Delta n_t)$ for $t \geq 2$ in Equation (13). Here, we also need a correcting factor $-|\Delta n_t|$. For two father nodes, each new node passes through this path four times. Using the methods mentioned above, we have

$$d(\Delta n_t, \Delta n_t) = 4g(t-2) + 2(|\Delta n_t| - 4)(|\Delta n_t| - 5) - |\Delta n_t| + d(n_0, \Delta n_t) + 4|\Delta n_t| \qquad (18)$$

Substituting Equations (1), (2), (17) and (18) into Equation (13), we have

$$g(t) = 3g(t-1) + 6g(t-2) - 8g(t-3) + 27 \times 2^{2t} - 18 \times 2^t - 40 + d(n_0, \Delta n_t) + 2d(n_0, n_{t-1}) - 2d(n_0, \Delta n_{t-1}) \qquad (19)$$

If $t$ is even, the number of the hops between any node from $\Delta n_t$ and the four early nodes is $t / 2$. If $t$ is odd, the number of the hops between any node from $\Delta n_t$ and the nodes $n_1$ is $(t - 1) / 2$. So, if $t$ is even, and we have

$$d(n_0, \Delta n_t) = 4 \times (t/2 + 3/4) \times \Delta n_t = (2t + 3) \times \Delta n_t = (2t + 3) \times 3 \times 2^t \qquad (20)$$

and

$$d(n_0, n_t) = d(n_0, n_0) + d(n_0, \Delta n_1) + d(n_0, \Delta n_2) + \cdots + d(n_0, \Delta n_t)$$

$$= 12 + (2 \times 1 + 3) \times 3 \times 2^1 + (2 \times 2 + 3) \times 3 \times 2^2 + \cdots + (2t + 3) \times 3 \times 2^t$$
$$= (12t + 6) \times 2^t + 6 \tag{21}$$

If $t$ is odd, and we have

$$d(n_0, \Delta n_t) = 4 \times ((t-1)/2 \times \Delta n_t + (\Delta n_t - 12)/6 \times 3 + (\Delta n_t - 12)/6 \times 3 \times 2 + 3 \times 3)$$
$$= 4 \times (t/2 + 1) \times \Delta n_t - 36 = (2t + 4) \times \Delta n_t - 36 = (2t + 4) \times 3 \times 2^t - 36 \tag{22}$$

and

$$d(n_0, n_t) = d(n_0, n_0) + d(n_0, \Delta n_1) + d(n_0, \Delta n_2) + \cdots + d(n_0, \Delta n_t)$$
$$= 12 + (2 \times 1 + 4) \times 3 \times 2^1 - 36 + (2 \times 2 + 4) \times 3 \times 2^2 - 36 + \cdots + (2t + 4) \times 3 \times 2^t - 36$$
$$= 12(t+1) \times 2^t - 36t \tag{23}$$

Substituting Equations (20), (21), and Equations (22), (23) into Equation (19) respectively, we obtain

$$g(t) = 3g(t-1) + 6g(t-2) - 8g(t-3) + 27 \times 2^{2t} + 12t \times 2^t - 18 \times 2^t - 28 \quad \text{if } t \text{ is even and } t \geq 3$$

and

$$g(t) = 3g(t-1) + 6g(t-2) - 8g(t-3) + 27 \times 2^{2t} + 12t \times 2^t - 12 \times 2^t - 72t + 68 \quad \text{if } t \text{ is odd and } t \geq 3.$$

For $g(0) = 12$, $g(1) = 150$ and $g(2) = 1014$, we hold

$$g(t) = ((648t + 557) \times 2^{2t}) + (11 - 324t) \times 2^t + 84t - 264) / 27 \text{ if } t \text{ is even} \tag{24}$$
$$g(t) = ((648t + 541) \times 2^{2t}) - (324t + 203) \times 2^t + 108t^2 + 336t - 96) / 27 \text{ if } t \text{ is odd} \tag{25}$$

Substituting Equations (24) and (25) into Equation (11), we obtain

$$L(t) = \begin{cases} \dfrac{[(648t + 580) \times 2^{2t} - (324t - 7) \times 2^t + 84t - 270]}{27(3 \times 2^{t+1} - 2)(3 \times 2^{t+1} - 1)} & \text{if } t \text{ is even} \\ \dfrac{[(648t + 544) \times 2^{2t} - (324t + 206) \times 2^t + 108t^2 + 336t - 102]}{27(3 \times 2^{t+1} - 2)(3 \times 2^{t+1} - 1)} & \text{if } t \text{ is odd} \end{cases} \tag{26}$$

For $t \to \infty$, we can hold

$$L(t) \approx 2t / 3 \tag{27}$$

From Equations (2) and (27), for $t \to \infty$, we can obtain the approximate analytical solution of the characteristic path length denoted by $L \approx 2 / (3\ln(2)) \ln(n) \approx 0.96\ln(n)$, and that of the network diameter $D \approx \ln(n) / \ln(2) \approx 1.46\ln(n)$. The network diameter is about 1.52 times that of the characteristic path length. Figure 3 shows the impact of the iteration steps and the network size on the analytical and simulation values of characteristic path length, and that of them on analytical value of network diameter. We see a straightforward dependency: the characteristic path length and the network diameter increases linearly with the iteration steps, the analytical value of characteristic path length is equal to the simulation value, and network diameter is no less than the value of characteristic path length or $\ln(n)$ in Figure 3a; the characteristic path length and the network diameter are both directly proportional to the logarithm of the network size in Figure 3b.

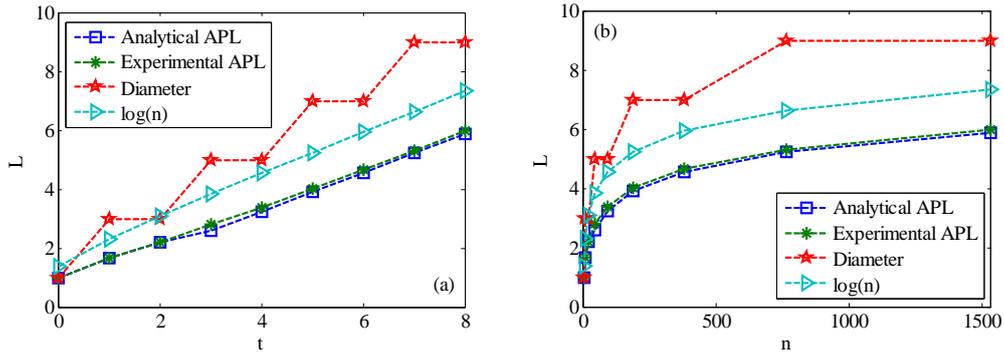

**Figure 3. The characteristic path length of the DMG model for small-world networks. (a)** The characteristic path length *L* and the diameter *D* versus the iterations steps *t*; **(b)** The characteristic path length *L* and the diameter *D* versus the networks size *n*

## 4 Conclusion and discussion

In conclusion, we have proposed a DMG model for small-world networks with geometric space structure in deterministic way. We have derived accurate analytical solutions of some main properties of the DMG model, and discussed them. More to the point, we have obtained an accurate analytical expression of characteristic path length. Numerical and experimental results show the accumulative degree is distributed exponentially and the network has high global clustering and small characteristic path length. The DMG model holds distinguishing small-world effect. In fact, many real networks have both small-world feature and geometric space structure and the DMG model can guide the research and the development of complex networks with geometric space structure, such as airport networks, metabolic pathways and social networks.

## Acknowledgments

This work was supported in part by the National Natural Science Foundation of China under Grants No. 60973129, the China Postdoctoral Science Foundation under Grant No. 200902324, the Guangdong Provincial Natural Science Foundation of China under Grant No. S2011010000812, the Program for Excellent Talents in Hunan Normal University of China under Grant No. ET10902, the Scientific Research Fund for Doctor Startup Project in Hunan Normal University under Grant No. 110608, and the Program for Changjiang Scholars and Innovative Research Team in University under Grant No. IRT0964.